\documentclass[12pt]{article}
\usepackage[latin1]{inputenc}

\usepackage{amsmath}
\usepackage{color}
\usepackage{amsfonts}
\usepackage{amssymb}
\usepackage{graphicx}
\usepackage{geometry}
\usepackage{amssymb,epsfig,subfigure}
\usepackage{hyperref}
\usepackage{comment}
\usepackage[font=footnotesize]{caption}

\usepackage[numbers,sort&compress]{natbib}

\def\simge{
    \mathrel{\rlap{\raise 0.511ex 
        \hbox{$>$}}{\lower 0.511ex \hbox{$\sim$}}}}

\def\simle{
    \mathrel{\rlap{\raise 0.511ex 
        \hbox{$<$}}{\lower 0.511ex \hbox{$\sim$}}}}

\makeatletter
\renewcommand\section{\@startsection {section}{1}{\z@}%
                                 {-3.5ex \@plus -1ex \@minus -.2ex}
                                   {2.3ex \@plus.2ex}%
                                   {\normalfont\large\bfseries}}
\renewcommand\subsection{\@startsection{subsection}{2}{\z@}%
                                   {-3.25ex\@plus -1ex \@minus -.2ex}%
                                     {1.5ex \@plus .2ex}%
                                     {\normalfont\bfseries}}
\renewcommand\subsubsection{\@startsection{subsubsection}{3}{\z@}%
                                   {-3.25ex\@plus -1ex \@minus -.2ex}%
                                     {1.5ex \@plus .2ex}%
                                     {\normalfont\itshape}}
\makeatother

\def\pplogo{\vbox{\kern-\headheight\kern -29pt
\halign{##&##\hfil\cr&{\ppnumber}\cr\rule{0pt}{2.5ex}&\ppdate\cr}}}
\makeatletter
\def\ps@firstpage{\ps@empty \def\@oddhead{\hss\pplogo}%
  \let\@evenhead\@oddhead 
}
\def\maketitle{\par
 \begingroup
 \def\thefootnote{\fnsymbol{footnote}}
 \def\@makefnmark{\hbox{$^{\@thefnmark}$\hss}}
 \if@twocolumn
 \twocolumn[\@maketitle]
 \else \newpage
 \global\@topnum\z@ \@maketitle \fi\thispagestyle{firstpage}\@thanks
 \endgroup
 \setcounter{footnote}{0}
 \let\maketitle\relax
 \let\@maketitle\relax
 \gdef\@thanks{}\gdef\@author{}\gdef\@title{}\let\thanks\relax}
\makeatother

\numberwithin{equation}{section}

\newcommand{\be}{\begin{equation}}
\newcommand{\bea}{\begin{eqnarray}}
\newcommand{\ee}{\end{equation}}
\newcommand{\eea}{\end{eqnarray}}
\newcommand\beq{\begin{equation}}
\newcommand\eeq{\end{equation}}

\newcommand{\mc}{\mathcal}

\renewcommand{\t}{\tilde}

\newcommand{\aU}{{\alpha_{UV}}}
\newcommand{\aI}{{\alpha_{IR}}}
\newcommand{\zU}{{z_{UV}}}
\newcommand{\zI}{{z_{IR}}}
\newcommand{\DU}{{\Delta_{UV}}}
\newcommand{\DI}{{\Delta_{IR}}}
\newcommand{\TT}{\langle \Theta(x) \Theta(0)\rangle}
\newcommand{\So}{{S_\text{on-shell}}}

\def\be{\begin{equation}}
\def\ee{\end{equation}}
\def\ba#1\ea{\begin{align}#1\end{align}}
\def\bg#1\eg{\begin{gather}#1\end{gather}}
\def\bm#1\em{\begin{multline}#1\end{multline}}
\def\bmd#1\emd{\begin{multlined}#1\end{multlined}}

\def\({\left(}
\def\){\right)}
\def\[{\left[}
\def\]{\right]}

\begin{document}

\setcounter{page}0
\def\ppnumber{\vbox{\baselineskip14pt
}}
\def\ppdate{
} \date{}

\author{Horacio Casini, Eduardo Test\'e, Gonzalo Torroba\\
[7mm] \\
{\normalsize \it Centro At\'omico Bariloche and CONICET}\\
{\normalsize \it S.C. de Bariloche, R\'io Negro, R8402AGP, Argentina}
}

\bigskip
\title{\bf  Holographic RG flows, entanglement entropy and the sum rule
\vskip 0.5cm}
\maketitle

\begin{abstract}
We calculate the two-point function of the trace of the stress tensor in holographic renormalization group flows between pairs of conformal field theories. We show that the term proportional to the momentum squared in this correlator gives the change of the central charge between fixed points in $d=2$ and in $d>2$ it gives the holographic entanglement entropy for a planar region. This can also be seen as a holographic realization of the Adler-Zee formula for the renormalization of Newton's constant. Holographic regularization is found to provide a perfect match of the finite and divergent terms of the sum rule, and it is analogous to the regularization of the entropy in terms of mutual information. Finally, we provide a general proof of reflection positivity in terms of stability of the dual bulk action, and discuss the relation between unitarity constraints, the null energy condition and regularity in the interior of the gravity solution.
\end{abstract}
\bigskip

\newpage

\tableofcontents

\vskip 1cm

\section{Introduction}\label{sec:intro}

In order to understand nonperturbative aspects of quantum field theories (QFT), it is of considerable interest to study renormalization group (RG) flows between pairs of conformal field theories CFT$_{UV}$ and CFT$_{IR}$. These RG flows are generically\footnote{Certain supersymmetric CFTs have moduli spaces of vacua, and it is then possible to have RG flows with spontaneous breaking of scale invariance.} triggered by turning on relevant operators $\mc O_i$ in the UV fixed point,
\be\label{eq:Sintro}
S=S_{UV}+ \int d^dx \,g_i \mc O_i(x)\,.
\ee
For flows that preserve Poincar\'e invariance (as will be the case in this work), the $\mc O_i$ are scalar operators with scaling dimension $\Delta_i<d$ at the UV fixed point.

These RG flows can be (partially) characterized by the correlators of the stress-tensor trace $\Theta(x) = T^\mu_\mu(x)$. One reason for this is that $\Theta(x)$ is not an independent operator of the theory, but rather is determined in terms of $\mc O_i$ and the $\beta$ functions of the couplings $g_i$ in (\ref{eq:Sintro}) via the operatorial relation $\Theta(x) = \beta_i \mc O_i(x)$ (up to a conformal anomaly function). The best understood case corresponds to flows between two-dimensional CFTs. Here unitarity of $\Theta(x)$ implies Zamolodchikov's c-theorem, and its two-point function yields the sum rule~\cite{Zamolodchikov:1986gt, cappelli}
\be\label{d2}
C_{UV}-C_{IR}=3 \pi\int d^{2}x\,\,x^2 \,\langle 0| \Theta(x)\Theta(0)| 0\rangle\,,
\ee
where $C_{UV}$ and $C_{IR}$ are the central charges of the UV and IR fixed points.

The situation in higher dimensions is more complicated and interesting. Early efforts were oriented at studying the stress-tensor two point function in $d>2$~\cite{cappelli,anselmi}; however, in general there is no clear connection of this quantity to global aspects of the RG. Instead, the generalization of  (\ref{d2}) to $d=4$ involves the 4-point function of $\Theta(x)$, and it has been shown that unitarity implies the $a$-theorem $a_{UV}>a_{IR}$~\cite{Komargodski:2011vj}.
Nevertheless, the question remains whether (and how) $\langle \Theta(x) \Theta(0) \rangle$ encodes some nontrivial properties of the RG flow. In fact, it turns out that this two-point function is related to two very different objects: the entanglement entropy (EE) for a planar surface, and the renormalization of Newton's constant for a background metric. Let us review how this connection comes about. 

For a planar entangling surface, rotational symmetry implies that the structure of the density matrix is surprisingly simple and universal. It is given by a thermal state with respect to boost ``time'' evolution, at a fixed dimensionless temperature $(2\pi)^{-1}$. 
Though this is an old result of axiomatic QFT \cite{BW}, only recently this fact has been used to provide general results for the EE of a planar surface in terms of correlation functions. Rosenhaus and Smolkin \cite{rs} proposed a simple way to compute the planar EE perturbing with relevant operators. 
In \cite{area} it was shown that following this route one arrives at a result that matches the Adler-Zee formula \cite{AZ} for the renormalization of Newton's constant. More concretely, for a large planar entangling surface of area $A_{\parallel}$, the entropy has the form 
\begin{equation}
\mc S=k \frac{A_{\parallel}}{\epsilon^{d-2}}+\mu\, A_{\parallel}\,,\label{dress}
\end{equation}
where $k$ is a non universal constant, $\epsilon$ is a short distance cutoff and $\mu$ is a constant of dimension $d-2$ that depends on the mass scales of the theory and may contain also non universal contributions.  The result of \cite{area} is the identification
\begin{equation}
\mu= -\frac{\pi}{d(d-1)(d-2)}\int_{|x|>\delta} d^{d}x\,\,x^2 \,\langle 0| \Theta(0)\Theta(x)| 0\rangle \,.\label{fgf}
\end{equation}
Here $\langle 0| \Theta(0)\Theta(x)| 0\rangle$ is the connected correlator evaluated in Euclidean space and the infinitesimal cutoff $\delta$ has just the purpose of eliminating contact terms. 

Eq.~(\ref{fgf}) is essentially the Adler-Zee formula \cite{AZ} for the renormalization of Newton's constant when quantum fields on a weakly curved background are integrated out. That is, we have\footnote{There are however exceptions to this identification between entanglement entropy and
 Newton's constant renormalization for theories with non unique stress tensors such as free scalars \cite{area}. In that case (\ref{fgf}) holds
  for a special (canonical) choice of stress tensor, and $\Delta (4G_N)^{-1}$ may contain additional terms due to couplings with the curvature.} 
\be
\mu=\Delta \left(\frac{1}{4 G_N} \right)\,.\label{iden}
\ee
 In fact $\mu$ in (\ref{dress}) can be interpreted as a dressing of the area term in the EE as we scale a region from small to large sizes. The same dressing occurs for black hole entropies as the black hole radius crosses the mass scales of the quantum fields, and (\ref{iden}) is the statement that the Bekenstein-Hawking entropy formula holds for large black holes independently of the matter content of the model. 
 
The identification of black hole entropy with entanglement entropy has a long history, starting with~\cite{sorkin}. Susskind and Uglum proposed that 
 entanglement entropy should renormalize in the same way as $(4G_N)^{-1}$~\cite{otro}. The subject was revisited several times in the past~\cite{revisited}.

In this paper we will not be concerned with Newton's constant renormalization, but rather focus on the formula (\ref{fgf}) for the area term in entanglement entropy in terms of stress tensor correlators. One problem with this relation is that both sides are very hard to evaluate in interacting theories. For this reason, we focus on CFTs and RG flows that admit a dual gravity description.
We will show that (\ref{fgf}) is satisfied holographically by explicitly computing both sides of the equation for any spacetime dimension $d$. Our main technical result is the computation of the two point correlator  $\langle 0| \Theta(0)\Theta(x)| 0\rangle$ for a general deformation of the ultraviolet (UV) CFT by a relevant perturbation. Then we will match the sum rule (\ref{fgf}) with the EE calculation in terms of minimal surfaces \cite{rt}. Previous holographic studies include \cite{holo,area-holo, f-theo}.

Another problem with (\ref{fgf}) is that in general both sides of the equation contain divergences. In particular, if the UV fixed point is perturbed with a relevant operator of dimension $\Delta\ge (d+2)/2$, the area term coefficient $\mu$ in EE calculated holographically diverges~\cite{holo}. The same counting follows from the right hand side of (\ref{fgf}) since $\langle 0| \Theta(0)\Theta(x)| 0\rangle\sim |x|^{-2\Delta}$ for short distances. When divergences are present, matching of both sides of (\ref{fgf}) for the divergent terms cannot be expected on general grounds. The universal part is the finite term or the logarithmic term in the case logarithmic terms are present; (\ref{fgf}) should then be understood as matching the universal parts. Notice the change in Newton's constant (\ref{fgf}), if finite, is negative, corresponding to antiscreening of gravity. If divergences appear the universal part can have positive sign.    

We will show that the standard holographic regularization given by a radial cutoff from the AdS boundary can be used to compute both sides of the equation giving a perfect match for the universal terms. They also coincide with the constant (or logarithmic) term in the mutual information between two parallel planes, as was argued in \cite{area} (see also \cite{F}). Moreover, our holographic sum rule will provide a unified description of the $d=2$ result, where the renormalization of the area term in EE is \cite{calcar}
\be\label{eq:2dsum}
\mu=-\frac{C_{UV}-C_{IR}}{6}\log(m\epsilon)\,,
\ee
(here $m$ is a mass scale for the RG flow) and the case $d>2$.

Finally, let us remark that the calculation of correlation functions for $\Theta(x)$ in holographic RG flows is formally very similar to the evaluation of scalar perturbations during cosmological inflation~\cite{Maldacena:2002vr}. This cosmological approach was recently applied to AdS/CFT in~\cite{Kaplan:2014dia}, who reproduced the sum rule for $d=2$. Our method in general dimension was motivated by this work, but differs significantly in the calculation of the stress tensor two-point function, as we explain below.

The paper is structured as follows. First, in \S \ref{sec:setup} we introduce the holographic setup and review some properties of holographic renormalization and the Hamiltonian approach that will be used in the paper. In \S\ref{sec:proof} we calculate the two-point function of $\Theta(x)$ for holographic RG flows between CFTs, and establish the sum rule (\ref{fgf}). Some consequences and applications are discussed in \S \ref{sec:appl}, including the relation to mutual information, properties of the stress-tensor spectral function, and a holographic proof of reflection positivity. Finally, \S \ref{sec:concl} contains our conclusions and various future directions motivated by the present results.

\section{The setup}\label{sec:setup}

We consider a renormalization group flow between a $d$-dimensional conformal field theory in the UV and a different CFT in the IR, triggered by turning on a relevant deformation,
\be\label{eq:Spert1}
S= S_{CFT}+ \int d^dx \, g \,\mc O(x)\,.
\ee
Here $\mc O$ is a scalar operator of CFT$_{UV}$ with conformal dimension $\DU<d$ and $g$ is a relevant, constant, coupling. At the endpoint of the flow, $\mc O$ becomes irrelevant, with dimension $\DI >d$ with respect to the infrared CFT.

The trace $\Theta(x)=T^\mu_\mu(x)$ of the energy-momentum tensor vanishes in the CFT, but becomes nontrivial due to the flow. Our goal is to calculate its two point function $\TT$. In particular, we want to evaluate
\be
\int d^dx\, x^2 \TT
\ee
and show that this gives the change in the central charge $C_{UV}-C_{IR}$ in $d=2$, eq. (\ref{d2}). For $d>2$, this should be proportional to the area term in the entanglement entropy of a large region~\cite{area}.

It is very hard to perform this explicit calculation in an interacting QFT. The computation of $\TT$ has been done for nearly free fields or in weakly coupled flows. Here we will use holography to obtain $\TT$ in strongly interacting RG flows that admit a gravity dual.

\subsection{Gravity description}\label{subsec:gravity}

A model for the gravity dual of the RG flow that we just described corresponds to a radial domain wall in $d+1$ dimensions that interpolates between an $AdS$ space with radius $L_{UV}$ when $r\to \infty$ and another $AdS$ with radius $L_{IR}$ when $r \to -\infty$. These endpoints of the domain wall are dual to CFT$_{UV}$ and CFT$_{IR}$ above. On the other hand, the relevant deformation of CFT$_{UV}$ by a scalar operator $\mc O$ means that the $d+1$-dimensional bulk solution is sourced by a scalar field that rolls on a nontrivial potential $V(\phi)$.

This holographic RG flow may be described by an euclidean action for Einstein-Hilbert gravity coupled to a scalar field,\footnote{We work in euclidean signature, and $\kappa^2=8\pi G^{(d+1)}$, where $G^{(d+1)}$ is Newton's constant in $d+1$ dimensions.}
\be\label{eq:S1}
S= \int d^{d+1}x \sqrt{g}\,\left( -\frac{1}{2\kappa^2} R^{(d+1)}+\frac{1}{2} g^{MN} \partial_M \phi \partial_N \phi+V(\phi)\right)\,.
\ee
The action has some additional boundary terms that will be discussed in \S \ref{subsec:review}.  It is possible to add higher derivative corrections or multiple fields but we restrict the analysis to this action for simplicity. We will comment on more general matter sectors in \S \ref{subsec:gen-matter}.
 
  We consider a potential that has a maximum at $\phi=0$ and admits an expansion
\be
V= V_{UV}+\frac{1}{2} m_{UV}^2 \phi^2+\ldots
\ee
 There is also a minimum at $\phi=\phi_0$,
\be
V=V_{IR}+\frac{1}{2} m_{IR}^2 (\phi-\phi_0)^2+\ldots
\ee

The domain-wall solution is described by
\be\label{eq:domain-g}
ds^2= dr^2+e^{2A(r)}  \delta_{\mu\nu} dx^\mu dx^\nu\;,\;\phi=\phi(r)\,.
\ee
The warp factor $A(r)$ and the scalar profile $\phi(r)$ satisfy Einstein's equations
\be\label{eq:GReqs}
\frac{1}{2\kappa^2}d(d-1) \dot A^2= \frac{1}{2}\dot \phi^2-V(\phi)\;,\;\frac{1}{\kappa^2}(d-1) \ddot A = -\dot \phi^2\,,
\ee
and the scalar field equation (which follows from the above)
\be
\ddot \phi+d\, \dot A \,\dot \phi - \partial_\phi V=0\,.
\ee
Dots denote derivatives with respect to $r$.

For $r\to \infty$ the domain wall starts near $\phi=0$ which, from these equations, gives an AdS solution with radius $L_{UV}$
\be\label{eq:AV}
A(r) \sim \frac{r}{L_{UV}}\;,\;-V_{UV}=\frac{d(d-1)}{2\kappa^2 L_{UV}^2}\,.
\ee
The endpoint of the wall occurs as $\phi$ reaches the minimum $\phi_0$, which corresponds in our coordinates to $r\to -\infty$ with
\be
A(r) \sim \frac{r}{L_{IR}}\;,\;-V_{IR}=\frac{d(d-1)}{2\kappa^2 L_{IR}^2}\,.
\ee

According to the AdS/CFT dictionary, the relation to the dimension $\DU$ of the dual operator $\mc O$ is 
\be  m_{UV}^2L_{UV}^2= \DU (\DU-d)\,.\ee
 Note that $m_{UV}^2<0$ since $\mc O$ is relevant. At the infrared we have analogously
 \be  m_{IR}^2L_{IR}^2= \DI (\DI-d)\,,\ee
with $\DI>d$ and $m_{IR}^2>0$.

We will not need the explicit domain wall profile for our calculation, but we can give more details about the behavior of $\phi(r)$ in the two asymptotic AdS regions. First we recall the solution for a massive scalar in AdS,
\be\label{eq:scalarAdS}
\phi(r) = \phi_0\, e^{-(d-\Delta)\frac{r}{L}}+\phi_\Delta\, e^{-\Delta \frac{r}{L}}\,.
\ee
We will restrict to a relevant perturbation in the range 
\be\label{eq:dim-restrict}
\Delta_{UV}>d/2\,,
\ee 
corresponding to the standard quantization.\footnote{For $\Delta_{UV}<d/2$, the alternate quantization has to be used. To our knowledge, holographic RG flows in this range are not fully understood yet.}
 In this case, the first term dominates at large $r$ and is dual to turning on a source $g$ in (\ref{eq:Spert1}). The second term is dual to the expectation value $\langle \mc O \rangle$. Since we are studying RG flows due to relevant deformations, $\phi_0\neq 0$ in the UV region of the domain wall. The domain wall is then described by an expansion of the form
\be
\phi(r) = e^{-(d-\DU)\frac{r}{L_{UV}}} \left(\phi^0_{UV}+\phi_\Delta e^{-(2\Delta_{UV}-d)\frac{r}{L_{UV}}}+\phi_2\, e^{-2\frac{r}{L_{UV}}}+\ldots \right)
\ee
at large $r$. 
On the other hand, in the IR region $r\to -\infty$ regularity requires that there is no term proportional to $e^{-\DI \frac{r}{L_{IR}}}$, and the profile is then of the form
\be
\phi(r) \approx \phi^0_{IR} \,e^{-(d-\DI)\frac{r}{L_{IR}}} \,.
\ee

\subsection{Holographic correlation functions}\label{subsec:review}

Before proceeding to the explicit calculation in the next section, it will be useful to review a few aspects of the holographic dictionary that we will need below. We will also recall the Hamiltonian form of the gravitational action, which will be useful in the computation.

In the semiclassical, large $N$ approximation, the AdS/CFT correspondence identifies the partition function of the QFT side with the on-shell action in the bulk, $\log Z_{QFT}= - \So$.
Correlation functions with $n$ points are obtained by turning on source terms for the dual bulk fields, computing the on-shell action and then taking $n$ derivatives with respect to the sources~\cite{Aharony:1999ti}. The stress-tensor trace couples to the trace of the boundary metric; this source is obtained by varying the warp factor of the domain wall (\ref{eq:domain-g}).
For the connected two-point function of the trace of the stress tensor, this gives
\be
\langle \Theta(x) \Theta(y)\rangle=-\frac{1}{\sqrt{h}} \frac{\delta}{\delta(\delta A_0(x))}\,\frac{1}{\sqrt{h}} \frac{\delta \So}{\delta(\delta A_0(y))}\,.
\ee
In more detail, the bulk metric gets perturbed with a boundary value $\delta A_0$,
\be
h_{\mu\nu}(x,r)=e^{2A(r)+2 \delta A(x,r)}\delta_{\mu \nu}\;,\;\lim_{r\rightarrow \infty}\delta A(x,r)=\delta A_0(x)\,.
\ee
At this order, we then need to solve the linearized bulk equations of motion allowing for a perturbation $\delta A_0(x)$.

There are three issues that complicate this calculation. First, unlike the graviton tensor mode --which is dual to the traceless part of the stress tensor, of protected dimension $d$-- the scalar metric mode mixes with fluctuations of the scalar field. Both are related by the constraint parts of Einstein's equation, resulting in a rather involved set of equations. From the perspective of the dual, this encodes the fact, noted above, that $\Theta$ is not an independent operator, but rather satisfies $\Theta(x) = \beta_g \mc O(x)$. A similar problem arises in inflationary perturbations, and we will find it useful to adapt some of the methods from cosmology to our situation.

The second problem regards how to solve the linearized equations in the bulk. These admit two arbitrary constants near the UV, as in (\ref{eq:scalarAdS}). The constant multiplying the subleading series (e.g. the `VEV' term $\phi_\Delta$ in (\ref{eq:scalarAdS})) is then fixed by requiring regularity as $r \to -\infty$. This is easy to implement in a pure AdS background, but this nonlocal differential problem becomes quite nontrivial in the presence of a domain wall. Indeed, we want to impose this regularity condition for any domain wall solution, so that we can make general statements regarding $\TT$. We will address this problem in \S \ref{sec:proof}, where we will find an analytic result for arbitrary flows in the limit of small momentum, as well as a series expansion for larger $p$.

Finally, the action (\ref{eq:S1}) diverges when evaluated on-shell, due to contributions from the UV AdS region. Fortunately, the solution to this issue is by now well understood using holographic renormalization~\cite{Skenderis:2002wp}. The method consists of making the on-shell action finite by adding terms that are covariant on the geometric quantities of the boundary. In our case, the action including the Gibbons-Hawking boundary term and the counterterms is
\be
S= \int d^{d+1}x \sqrt{g}\,\left( -\frac{1}{2\kappa^2} R^{(d+1)}+\frac{1}{2} g^{MN} \partial_M \phi \partial_N \phi+V(\phi)\right)-\frac{1}{\kappa^2} \int d^dx \sqrt{h}\,K+S_{ct}\,.
\ee
Here $K$ is the trace of the extrinsic curvature of the boundary  metric (discussed in more detail below), and
\be\label{eq:Sct}
S_{ct}=\frac{d-1}{\kappa^2} \int d^d x \sqrt{h} \left(\frac{1}{L_{UV}}+\frac{L_{UV}}{2(d-1)(d-2)}R^{(d)}+\frac{\kappa^2}{2}\frac{d-\Delta}{d-1}\phi^2+\ldots \right)\,.
\ee
The first two counterterms were found in~\cite{Balasubramanian:1999re} by requiring a finite energy-momentum tensor; the one proportional to $\phi^2$ cancels the boundary term generated when integrating by parts to evaluate the scalar field action on-shell.

\subsection{Hamiltonian formulation}\label{subsec:hamiltonian}

In order to compute the action to quadratic order, it will be convenient to use the Hamiltonian form of the Einstein-Hilbert action~\cite{Arnowitt:1962hi}. The reason is that various aspects of the holographic RG simplify in the Hamiltonian approach, as found in~\cite{Akhmedov:1998vf, deBoer:1999xf, Akhmedov:2002gq, Papadimitriou:2004ap, Papadimitriou:2004rz}, and more recently in~\cite{Heemskerk:2010hk, Faulkner:2010jy, Dong:2012afa}.\footnote{Here we follow the conventions in~\cite{Fukuma:2000bz}.}

One begins from the ADM decomposition along the radial direction
\be\label{eq:ADM1}
ds^2= N(x,r)^2 dr^2+ h_{\mu\nu}(x,r) (dx^\mu+N^\mu(x,r) dr)(dx^\nu+N^\nu(x,r) dr)\,,
\ee
and the extrinsic curvature of an $r=\text{const}$ surface is given by
\be
K_{\mu\nu}=\frac{1}{2N}(\dot h_{\mu\nu}- \nabla_\mu N_\nu-\nabla_\nu N_\mu)\,.
\ee
Dots denote radial derivatives, $\nabla_\mu$ is the covariant derivative with respect to $h_{\mu\nu}$, and $K = h^{\mu\nu}K_{\mu\nu}$.

The action $S=S_{grav}+S_{matter}+S_{ct}$ in terms of the ADM variables becomes
\bea\label{eq:action-ADM}
S_{grav}&=&-\frac{1}{2\kappa^2} \int dr d^d x \sqrt{h}N \left(R^{(d)}+K^2-K_{\mu\nu}K^{\mu\nu} \right)\,, \\
S_{matter}&=& \int dr d^d x \sqrt{h}N\left(\frac{1}{2N^2}(\dot \phi-N^\mu \partial_\mu\phi)^2+\frac{1}{2} h^{\mu\nu} \partial_\mu \phi \partial_\nu \phi+V(\phi) \right)\nonumber\,.
\eea
The Gibbons-Hawking boundary term cancels when writing the $d+1$-dimensional curvature scalar in terms of $d$-dimensional quantities (see e.g.~\cite{poisson}). In first order form, where both the variable and its canonical momentum are treated as independent, the action reads
\be\label{eq:action-first}
S=\int dr d^d x \sqrt{h} \left(\frac{1}{2\kappa^2} \Pi^{\mu\nu} \dot h_{\mu\nu}+\Pi_\phi \dot \phi+N \mathcal H+N_\mu P^\mu \right)+S_{ct}\,,
\ee
with
\bea\label{eq:constraints}
\mathcal H&=&\frac{1}{2\kappa^2} \left(\frac{1}{d-1}(\Pi_\mu^\mu)^2-\Pi_{\mu\nu}^2 \right)-\frac{1}{2} \Pi_\phi^2+V(\phi)-\frac{1}{2\kappa^2} R^{(d)}+\frac{1}{2}h^{\mu\nu} \partial_\mu \phi \partial_\nu \phi \,,\nonumber\\
P^\mu&=&\frac{1}{\kappa^2} \nabla_\nu \Pi^{\mu\nu}-\Pi_\phi \nabla^\mu \phi\,.
\eea
The fields $N$ and $N_\mu$ are Lagrange multipliers, imposing the constraints
\be
\frac{1}{\sqrt{h}}\frac{\delta S}{\delta N}= \mathcal H =0\;,\;\frac{1}{\sqrt{h}}\frac{\delta S}{\delta N^\mu}= P_\mu =0\,.
\ee
Furthermore, the equations of motion for $\Pi_{\mu\nu}$ and $\Pi_\phi$ give the relations
\be\label{eq:momenta}
\Pi_{\mu\nu}= K_{\mu\nu}-h_{\mu\nu}K\;,\;\Pi_\phi=\frac{1}{N} \left(\dot \phi - N^\mu \partial_\mu \phi\right)\,,
\ee
which reproduce the momenta computed from (\ref{eq:action-ADM}).

\section{The stress-tensor two-point function}\label{sec:proof}

This section presents the main technical result of the paper: the calculation of $\TT$. We proceed in three steps. First we determine in \S \ref{subsec:quadratic} the action for the scalar metric fluctuation to second order. Next, in \S \ref{subsec:matching} we show how to solve the corresponding equation of motion imposing the regularity condition in the IR through a matching procedure. Finally, we compute the two-point function in a perturbative expansion around large distances in \S \ref{subsec:TT}. We end the section by establishing the holographic sum rule in \S \ref{subsec:holosum}.

\subsection{Quadratic action for the Weyl mode}\label{subsec:quadratic}

In order to compute $\TT$, we have to turn on a space-time dependent fluctuation of the metric, $h_{\mu\nu}(x,r) = e^{2A(r)+2 \delta A(x,r)} \delta_{\mu \nu}$, and then we need to evaluate the action on-shell to quadratic order in the fluctuation $\delta A$.

Without a convenient gauge choice, Einstein's equations lead to a complicated differential system that mixes $\delta A$ and $\delta \phi$. This is in part due to the constraints $\delta G_{\mu r}= \delta T_{\mu r}$ and $\delta G_{00}= \delta T_{00}$ that relate both modes. One possibility would be to work in terms of gauge invariant variables; however, we find it more convenient to work in the gauge
\be
h_{\mu\nu}(x,r)=e^{2A(r)+2 \delta A(x,r)}\delta_{\mu \nu}\;,\;\phi(x,r)=\phi(r)
\ee
so that all the fluctuations of the scalar field vanish. As shown in the similar problem of scalar perturbations during inflation, the equations simplify considerably with this choice~\cite{Maldacena:2002vr}.
Note that in this gauge, $N$ and $N_\mu$ in (\ref{eq:ADM1}) will become nontrivial. This gauge was also recently used in a related holographic setup in~\cite{Kaplan:2014dia}, which inspired our approach. As we note below, however, we differ from this work in important aspects of the analysis.

The quadratic action for $\delta A$ only requires $N$ and $N_\mu$ to first order in $\delta A$, because the second order terms appear multiplying the constraints $\mathcal H$ and $P^\mu$ evaluated on the background, which vanish since we work on a solution. At first order, we work with the ansatz
\be
N= 1+\delta N\;,\; N_\mu= e^{2A(r)} \partial_\mu \delta \psi\,,
\ee
which we will see solves the constraints. In this case,
\bea\label{eq:extrinsic}
K_{\mu\nu}&=&\frac{1}{N} \left((\dot A+\dot{\delta A}) h_{\mu\nu}-e^{2A(r)} \partial_\mu \partial_\nu\delta \psi \right)\,, \nonumber\\
K&=&\frac{1}{N} \left(d(\dot A+\dot{\delta A})- \Box \delta \psi\right)\,,
\eea
where $\Box f \equiv \delta^{\mu\nu} \partial_\mu \partial_\nu f$.

Consider first the momentum constraint, $\nabla^\mu \Pi_{\mu\nu}=0$. From (\ref{eq:momenta}) and (\ref{eq:extrinsic}), we obtain
\be\label{eq:deltaN}
\delta N = \frac{\dot{\delta A}}{\dot A}\,.
\ee
The solution for the Hamiltonian constraint $\mathcal H=0$ is more involved. First we evaluate the scalar curvature for $h_{\mu\nu}$:
\be
R^{(d)}=- (d-1)e^{-2(A+\delta A)} \left(2 \Box \delta A+(d-2) \delta^{\mu\nu}\partial_\mu \delta A\, \partial_\nu \delta A\right)\,,
\ee
which is valid to all orders in $\delta $. Plugging then this result and (\ref{eq:extrinsic}) into (\ref{eq:constraints}) obtains
\bea
\mathcal H& =& \frac{d-1}{2\kappa^2} \frac{1}{N^2} \left(d(\dot A+\dot{\delta A})^2-2(\dot A+\dot{\delta A}) \Box \delta \psi\right)\nonumber\\
&+&\frac{d-1}{2\kappa^2}e^{-2(A+\delta A)} \left(2 \Box \delta A+(d-2)( \partial_\mu \delta A)^2\right)-\frac{1}{2N^2} \dot \phi^2+V(\phi)\,.
\eea

As a check, the zeroth order in the fluctuation,
\be
\mathcal H^{(0)}=\frac{1}{2\kappa^2}d(d-1) \dot A^2-\frac{1}{2}\dot \phi^2+V(\phi)
\ee
reproduces the classical equation of motion (\ref{eq:GReqs}). Expanding next to first order in fluctuations obtains an equation that determines $\Box \delta \psi$,
\be\label{eq:psi}
\Box \delta \psi=-\frac{\ddot A}{\dot A} \dot{\delta A}+e^{-2 A} \frac{\Box \delta A}{\dot A}+O(\delta^2)\,,
\ee
where we used the value of $\delta N$ in (\ref{eq:deltaN}), and eliminated $\dot \phi^2$ in favor of $\ddot A$ using (\ref{eq:GReqs}).

We now plug (\ref{eq:deltaN}) and (\ref{eq:psi}) into (\ref{eq:action-first}) and expand to quadratic order in $\delta A$. Notice that, to this order, $N_\mu P^\mu=0$, and $N \mathcal H = \mathcal H^{(2)}$. After integration by parts, the terms $S_{grav}+S_{matter}$ of the action expanded to quadratic order can be brought to the form
\be\label{eq:S2}
S^{(2)}=-\frac{d-1}{2\kappa^2} \int dr d^dx \left[ e^{dA}\frac{\ddot A}{\dot A^2}\left(\dot{\delta A}^2+e^{-2A}(\partial_\mu \delta A)^2 \right)+ \frac{d}{dr}\left(\frac{e^{(d-2)A}}{\dot A}(\partial_\mu \delta A)^2+d^2 e^{dA} \dot A (\delta A)^2 \right)\right]\,.
\ee

We also need to include the counterterms (\ref{eq:Sct}) from holographic renormalization. Expanding $S_{ct}$ to quadratic order gives a contribution that cancels the boundary terms in (\ref{eq:S2}),\footnote{In particular, the first two terms of the $\dot A$ expansion near the boundary cancel the first and third terms in the counterterm action. This continues to higher orders.} so the final result for the quadratic action is
\be\label{eq:Squadfinal}
S=\frac{d-1}{2\kappa^2} \int dr d^dx \, e^{dA}\,\varepsilon(r)\,\left(\dot{\delta A}^2+e^{-2A}(\partial_\mu \delta A)^2 \right)\,,
\ee
where we have defined
\be
\varepsilon(r) \equiv -\frac{\ddot A}{\dot A^2}\,.
\ee

Therefore, transforming to Fourier modes, we need to solve the equation of motion
\be
\frac{d}{dr} \left(e^{dA}\varepsilon(r) \frac{d \delta A}{dr} \right)-e^{(d-2)A}\varepsilon(r) \, p^2 \, \delta A=0
\ee
with the boundary condition
\be
\delta A(p, r_{UV}) = \delta A_0(p)
\ee
and then compute the second derivative of the on-shell action with respect to $\delta A_0$. Evaluated on the equation of motion, only the term from integrating by parts in (\ref{eq:Squadfinal}) survives, and thus
\be
\So=\frac{d-1}{2\kappa^2}\int  d^dx \, e^{dA}\,\varepsilon(r)\,\delta A\,\partial_r \delta A\Big|_{r\to \infty}\,.
\ee

\subsection{Matching and solution}\label{subsec:matching}

It is now convenient to work with the conformal radial coordinate $z\in (0,\infty)$,
\be
dr=-a(z) dz\;,\;e^{A(r)}=a(z)\,,
\ee
in terms of which
\be
S=\frac{d-1}{2\kappa^2} \int d^dx\, dz \,a^{d-1}(z) \varepsilon(z) ((\partial_z \delta A)^2+ (\partial_\mu\delta A)^2)\,,
\ee
and
\be\label{eq:eomz}
\frac{d}{dz} \left(a^{d-1}(z) \varepsilon(z) \frac{d \delta A}{dz} \right)-a^{d-1}(z)\varepsilon(z) p^2 \delta A(z)=0\,.
\ee

The radial flow starts in the UV due to a source for a relevant operator or, in gravity, language,
\be
\lim_{z\to 0}\phi(z) \approx \phi^0_{UV} z^{d-\Delta_{UV}}\,,
\ee
with $\Delta_{UV}<d$. We also take $\Delta_{UV}>d/2$ to avoid subtleties with the alternate quantization. In the IR this flows to an irrelevant operator of dimension $\Delta_{IR}>d$, and
\be\label{eq:phiIR}
\lim_{z\to \infty}\phi(z) \approx \phi^0_{IR} z^{-(\Delta_{IR}-d)}\,.
\ee
Regularity in the IR requires that there is no mode proportional to $z^{\Delta_{IR}}$. We take the UV approximation to be valid for $z\lesssim \zU$, and the IR approximation good for $z\gtrsim \zI$. We will also treat $\zU$ as a UV regulator, sending $z_{UV} \to 0$ after appropriate subtraction of divergences. On the other hand, it is important that $z_{IR}$, although much larger than the mass scale of the dual RG flow, is finite.

We need to obtain $\varepsilon(z)$ near the UV and IR regions. Close to the AdS regions, the background equations of motion (\ref{eq:GReqs}) give
\be
\varepsilon(z)\approx -\frac{d}{2V} \dot \phi^2\,.
\ee
Using 
\be
\dot \phi \approx - \frac{d-\Delta}{L} \phi_0 z^{d-\Delta}\,,
\ee
and recalling the relation (\ref{eq:AV}) between $V$ and the AdS radius, obtains
\be\label{eq:varep}
\varepsilon(z) \approx \eta z^{2(d-\Delta)}\;,\;\eta \equiv (\kappa \phi_0)^2 \frac{(d-\Delta)^2}{d-1}\,.
\ee
For the warp factor, it is enough to retain the leading AdS behavior, $a(z)\approx L/z$.

We note here one of the main differences with~\cite{Kaplan:2014dia}. That work approximated $\varepsilon \approx \varepsilon_0$ in the UV and IR regions, taking $\varepsilon_0 \to 0$ at the end. From (\ref{eq:varep}), this corresponds to the limit $\Delta_{UV, IR} \to d$. Therefore, that approach only applies to a flow triggered by an almost marginal operator. Here we do not wish to impose this restriction, and hence we will use (\ref{eq:varep}) instead. In fact, we will find that the $z$ dependence in (\ref{eq:varep}) has important consequences for establishing the holographic sum rule.

We can now solve (\ref{eq:eomz}) in the asymptotic regions. In the UV and IR AdS regions,
\be
(z^{1-2\alpha}\delta A')'-p^2 z^{1-2\alpha} \delta A=0\,,
\ee
where primes denote derivatives with respect to $z$, and we have defined
\be
\alpha \equiv \Delta-\frac{d}{2}\,.
\ee
The general solution is of the form
\be
\delta A= (pz)^\alpha \left(c_1 I_\alpha(p z) +c_2 K_\alpha (pz) \right)\,.
\ee
Note that $\alpha>0$ in the UV region due to (\ref{eq:dim-restrict}); $\alpha$ is also positive in the IR, because the operator becomes irrelevant as the flow approaches the fixed point.
In the IR only $K_\alpha$ is regular. We then have
\bea\label{eq:solAdS}
\delta A_{UV}(z) &=& (pz)^{\aU}\left(\frac{2^{1-\aU}}{\Gamma(\aU)} h_0(p) K_{\aU}(pz) + 2^\aU \Gamma(1+\aU) h_1(p) I_\aU(pz)
\right)\,,\nonumber\\
\delta A_{IR}(z) &=&D_1(p) \,(p z)^{\aI}  \,K_\aI(pz)\,,
\eea
with arbitrary momentum-dependent factors $h_0$, $h_1$ and $D_1$.
Here $h_0$ is the boundary source for $\Theta$, and the goal is to determine $h_1/h_0$. We note for future use the expansions for small $p z$ in both limits,
\bea
\delta A_{UV}(z) &=& h_0(p) +  \left(\frac{\Gamma(-\aU)}{4^\aU \Gamma(\aU)} h_0(p) +h_1(p) \right)(p z )^{2\aU}+\ldots\nonumber\\
\delta A_{IR}(z) &=& \frac{\Gamma(\aI)}{2^{1-\aI}}D_1(p)+  \frac{\Gamma(-\aI)}{2^{1+\aI}}\,D_1(p)\,(p z )^{2\aI}+\ldots
\eea

It is in general not possible to find an analytic solution for general momentum $p$.\footnote{For some exact solutions in specific microscopic models see for instance~\cite{Bianchi:2001de}.} However, note that in order to evaluate (\ref{fgf}) we only require the correlator for small momentum up to order $p^2$. This will imply a great simplification in what follows, and it motivates looking for a solution in a perturbative expansion around $p=0$. 

For $p=0$ we have the exact solution
\be
\delta A_{p=0}(z) = A_2 +A_1 \int^z \frac{dz'}{a^{d-1}(z') \varepsilon(z')}\,,
\ee
which we use to construct a solution in powers of $p^2$,
\be\label{eq:solpert}
\delta A_{pert}(z)= A_2 (1+p^2 g_1(z) +\ldots)+ A_1 (f_0(z)+p^2 f_1(z) +\ldots)\,.
\ee
We have defined
\be\label{eq:f}
f_0(z)= \int_{z_{IR}}^z\,\frac{dy}{a^{d-1}(y) \varepsilon(y)}\;,\;f_1(z)=\int_{z_{IR}}^z\,\frac{dy_1}{a^{d-1}(y_1)  \varepsilon(y_1)} \int_{\zI}^{y_1} dy_2 a^{d-1}(y_2) \varepsilon(y_2) f_0(y_2)
\ee
and
\be\label{eq:g}
g_1(z)=\int_{z_{IR}}^z\,\frac{dy_1}{a^{d-1}(y_1)  \varepsilon(y_1)} \int_{\zI}^{y_1} dy_2 a^{d-1}(y_2) \varepsilon(y_2)\,.
\ee
Higher powers in $p^2$ can be obtained recursively,
\be
\left(\begin{matrix}f_n(z) \\ g_n(z)\end{matrix} \right)=\int_{z_{IR}}^z\,\frac{dy_1}{a^{d-1}(y_1)  \varepsilon(y_1)} \int_{\zI}^{y_1} dy_2 a^{d-1}(y_2) \varepsilon(y_2) \,\left(\begin{matrix}f_{n-1}(y_2) \\ g_{n-1}(y_2)\end{matrix}\right)\,.
\ee

The solution over all $z$ can be found when the above expansions overlap. This happens at small enough momentum, $p\, \zU \ll 1$ and $p\, \zI \ll 1$.\footnote{At the end of the calculation $\zU\to 0$, so $p \zU \ll 1$ is straightforward. On the other hand, $\zI$ is a finite radial scale; for a given fixed $\zI$ we have to choose momenta $p \zI \ll 1$.} In this regime we match (\ref{eq:solpert}) with (\ref{eq:solAdS}) and then obtain the consequence of the IR regularity condition on the UV expansion. This matching procedure was introduced in~\cite{Hoyos:2012xc}; see also~\cite{Bajc:2013wha}. We start from the IR. Note that we have defined all the integrals $f_i$ and $g_i$ in (\ref{eq:solpert}) to vanish at $z=\zI$. Therefore, matching the two solutions and their derivatives,
\bea
\delta A_{pert}(\zI) &=& A_2 = \delta A_{IR}(\zI) \,,\nonumber\\
\delta A_{pert}'(\zI)&=&  f_0'(\zI)\,A_1= \delta A_{IR}'(\zI)\,,
\eea
and hence
\be\label{eq:A1A2}
\frac{A_1}{A_2}= \frac{1}{f_0'(\zI)}\,\frac{\delta A_{IR}'(\zI)}{\delta A_{IR}(\zI)}\,.
\ee
Repeating the same procedure in the UV obtains
\bea
(f_0'+p^2 f_1'+\ldots) A_1+ (p^2 g_1'+\ldots) A_2&=& \delta A_{UV}'\,,\\
(f_0+p^2 f_1+\ldots) A_1+ (1+p^2 g_1+\ldots) A_2&=&\delta A_{UV}\,,\nonumber
\eea
and all functions are evaluated at $z=\zU$. Therefore,
\be\label{eq:ratioAUV}
\frac{\delta A_{UV}'}{\delta A_{UV}}\Big|_{z=\zU}=\frac{(f_0'+p^2 f_1'+\ldots) (A_1/A_2)+(p^2 g_1'+\ldots)}{(f_0+p^2 f_1+\ldots)(A_1/A_2)+ (1+p^2 g_1+\ldots)}\Big|_{z=\zU}\,.
\ee
with $A_1/A_2$ given by (\ref{eq:A1A2}).

In summary, for a given boundary value $h_0$, we find a unique solution in a series expansion at small momenta, and this solution is regular in the IR. The ratio $h_1/h_0$ is determined from (\ref{eq:ratioAUV}).

\subsection{Calculation of the stress tensor correlator}\label{subsec:TT}

We are now ready to compute $\langle \Theta(x)\Theta(0)\rangle$. For the connected correlator\footnote{If $\mc O$ has an expectation value, there is an additional disconnected contribution that appears as a term linear in $h_0$.} we need the quadratic term in the source $h_0$:
\be
S_\text{on-shell}=-\frac{d-1}{2\kappa^2} \int \frac{d^dp}{(2\pi)^d}\,a^{d-1}(z_{UV})\varepsilon(z_{UV})\,  \frac{\delta A_{UV}'(z_{UV})}{\delta A_{UV}(z_{UV})}\,h_0(p) h_0(-p)\Big|_{\zU \to 0}\,.
\ee
Then,\footnote{We are using the standard notation $\langle \Theta(p)\Theta(-p)\rangle=\int d^dx\, e^{i p x} \langle \Theta(0)\Theta(x)\rangle$.}
\be
\langle \Theta(p)\Theta(-p)\rangle =\frac{d-1}{\kappa^2}\,a^{d-1}(z_{UV})\varepsilon(z_{UV})\frac{\delta A_{UV}'}{\delta A_{UV}}\Big|_{\zU \to 0}\label{erase}
\ee
and this is the quantity that we obtain from the matching solution (\ref{eq:ratioAUV}). Noting that $a^{d-1}(z)\varepsilon(z)=1/f_0'(z)$, we arrive to
\be\label{eq:thetafinal}
\langle \Theta(p)\Theta(-p)\rangle=\frac{d-1}{\kappa^2}\frac{(1+p^2 f_1'/f_0'+\ldots) (A_1/A_2)+(p^2 g_1'/f_0'+\ldots)}{(f_0+p^2 f_1+\ldots)(A_1/A_2)+ (1+p^2 g_1+\ldots)}\Big|_{z=\zU}\,.
\ee
This is our final expression for the correlator of $\Theta(p)$ at small momentum, and is the main technical result of the paper.

In order to understand the momentum dependence of this correlator, we expand (\ref{eq:A1A2}) for small $p\,\zI$, finding
\be
\frac{A_1}{A_2} \propto p^{2\Delta_{IR}-d}\,.
\ee
Therefore, (\ref{eq:thetafinal}) contains terms that are nonanalytic in momentum (for generic $\Delta_{IR}$) of the form $p^{2\Delta_{IR}-d}(1+p^2+\ldots)$, together with terms that are analytic in $p^2$. Let us focus on the nonanalytic piece first,
\be\label{eq:TTIR}
\langle \Theta(p)\Theta(-p)\rangle=-\frac{1}{2^{2\aI}}\frac{\Gamma(1-\aI)}{\Gamma(\aI)}\,\left((d-\DI)L_{IR}^{\frac{d-1}{2}}\phi_{IR}^0 \right)^2\,p^{2\DI-d}+\ldots
\ee
Here $\phi_{IR}^0$ is given in terms of the domain wall scalar $\phi(z) \approx \phi_{IR}^0 z^{-(\DI-d)}$ at large $z$. This behavior matches the prediction from the operatorial relation $\Theta(x) = \beta_g \mc O(x)$ for a perturbation of the fixed point by a term in the action $\int d^dx\, g \mc O(x)$, where $\Delta(\mc O) = \Delta_{IR}$. Indeed, identifying the coupling with the holographic source, $g=L_{IR}^{\frac{d-1}{2}}\phi_{IR}^0$, the classical $\beta$ function is $\beta_g \approx (\DI-d)g$, and hence
\be
\langle \Theta(p)\Theta(-p)\rangle= \beta_g^2\,\langle \mc O(p)\mc O(-p)\rangle\,.
\ee
So our result is in agreement with the dual CFT answer. In the opposite limit of large momentum $p \zI \gg 1$, the perturbative problem is determined purely in terms of UV data: the solution is dominated by the $h_0$ term and no matching is needed up to exponentially small corrections from $h_1$. In this case we find (\ref{eq:TTIR}) with the replacement $\aI \to \aU$, in agreement again with the operator relation $\Theta(x) = \beta_g \mc O(x)$ near the UV fixed point.

Let us now focus on the analytic terms. At the UV fixed point the contributions analytic in $p^2$ are contact terms and hence depend on the regularization scheme; in our calculation we have chosen a specific regularization in terms of the holographic RG prescription described before. However, having fixed the scheme at the UV, the analytic terms become physical in the IR, and depend on global properties of the RG, which we now explore. 

At small momenta, the nonanalytic contributions from $A_1/A_2$ are subleading compared to $p^2$, because $\Delta_{IR}>d$. At leading order in $p^2$ we then obtain
\be\label{eq:temp2}
a^{d-1}(z_{UV})\varepsilon(z_{UV})\frac{\delta A_{UV}'}{\delta A_{UV}}\approx\,p^2\,\frac{g_1'(\zU)}{f_0'(\zU)}=p^2\,\int_{\zI}^{\zU} dz \,a^{d-1}(z) \varepsilon(z)\,.
\ee
From the point of view of the matching procedure, the $p^2$ term is then dominated by the first perturbative correction given by $g_1(z)$ in (\ref{eq:solpert}). This is another point where we differ from~\cite{Kaplan:2014dia}, who focused on the $p^0$ term.\footnote{Similar issues were identified in other contexts by~\cite{Bajc:2013wha}.}
Taking this into account obtains
\be\label{eq:TTsemifinal}
\int d^d x \,x^2 \langle \Theta(x) \Theta(0) \rangle =\frac{2d(d-1)}{\kappa^2} \int_0^\infty dz\, a^{d-1}(z) \varepsilon(z)\,,
\ee
where the factor of $2d$ comes from $-\nabla_p^2$, the Fourier transform of $x^2$, applied to $p^2$.
In terms of the $r$ variable introduced before,
\be
\int d^d x \,x^2 \langle \Theta(x) \Theta(0) \rangle =\frac{2d(d-1)}{\kappa^2} \int dr\,e^{(d-2) A(r)}\left(-\frac{\ddot A(r)}{\dot A(r)^2} \right)\,.
\ee
Integrating by parts, we arrive to
\be\label{eq:TTfinal}
\int d^d x \,x^2 \langle \Theta(x) \Theta(0) \rangle =\frac{2d(d-1)}{\kappa^2} \frac{e^{(d-2)A(r)}}{\dot A(r)} \Big|_{r_{IR}}^{r_{UV}}-\frac{2d(d-1)(d-2)}{\kappa^2} \int dr\, e^{(d-2) A(r)}\,.
\ee

\subsection{The holographic sum rule}\label{subsec:holosum}

Finally we are ready to establish the holographic sum rule.
For $d=2$, (\ref{eq:TTfinal}) gives the c-theorem,
\be
\int d^2 x \,x^2 \langle \Theta(x) \Theta(0) \rangle =\frac{4}{\kappa^2} \frac{1}{\dot A(r)} \Big|_{r_{IR}}^{r_{UV}}=\frac{4}{\kappa^2}(L_{UV}-L_{IR})=\frac{1}{3\pi}(C_{UV}-C_{IR})\,,
\ee
where in the last step we used the standard $d=2$ holographic relation $C=(3/2)(L/G)$.

For $d>2$, the first term in (\ref{eq:TTfinal}) is a UV divergence, while the second term is proportional to the holographic entanglement entropy for a planar entangling surface. This entropy is given by $ {\cal A}_{\textrm{bulk}}/(4 G^{(d+1)})$, with ${\cal A}_{\textrm{bulk}}$ the area of a bulk $(d-1)$-dimensional minimal surface anchored in the $(d-2)$-dimensional entangling surface in the boundary. For a planar entangling surface the bulk minimal surface extends right in the $r$ direction, and the entropy is
\be
\mc S=\frac{A_\parallel}{4 G^{(d+1)}} \int dr\, e^{(d-2) A(r)}\,.  
\ee
Using $\kappa^2=8\pi G^{(d+1)}$ in (\ref{eq:TTfinal}) we have
\be
-\frac{\pi}{d(d-1)(d-2)}\int d^d x \,x^2 \langle \Theta(x) \Theta(0) \rangle =\frac{\mc S}{A_{\parallel}}-\frac{1}{4 G^{(d+1)}(d-2)} \frac{e^{(d-2)A(r)}}{\dot A(r)} \Big|_{r=r_{UV}}\,,\label{formu}
\ee
giving a holographic realization of (\ref{fgf}). 

The second term on the right hand side gives a divergent boundary piece which exactly cancels the leading divergent term in the area. This is necessary for consistency, since for $\Delta<(d+2)/2$ the left hand side of (\ref{formu}) is finite, while the area is finite in this case once the leading divergence is subtracted. The universal constant term does not get corrected from this boundary term which only contains fractional powers of $z$ for generic $\Delta$. Powers of $z$ in the boundary term do not correct a logarithmic term when this is present in the entropy. In this case the constant term does get corrected, but is not universal. 

It is interesting to note that in this particular holographic cutoff given by $z_{UV}$ even the divergent terms match between both sides of (\ref{formu}), and the match of divergent terms in the entropy and the ones in the correlation function get corrected in a unique way by the boundary term. Furthermore, the holographic formula (\ref{eq:TTsemifinal}) provides a unified answer for the $d=2$ c-theorem and the area theorem in $d>2$.

\section{Applications}\label{sec:appl}

In this section we explore some of the physical consequences and applications of the holographic sum rule (\ref{eq:TTfinal}), (\ref{formu}). In order to understand better the role of the holographic regulator, in \S \ref{subsec:mutual} we compare the result from holographic regularization to the answer in terms of the mutual information, which introduces a point-splitting regularization. We next focus in \S \ref{subsec:positivity} on how unitarity --or its euclidean version, reflection positivity-- of the boundary theory is encoded in the bulk. We will show that in the large N limit reflection positivity is equivalent to stability of the gravitational action. We apply this to the spectral density for $\Theta(x)$, and show how the NEC and regularity of the solution give a unitary result. Motivated by possible relations to anomalies, \S \ref{subsec:pd} explores the structure of the $p^d$ term in the holographic stress tensor correlator, which is scale invariant. We end in \S \ref{subsec:gen-matter} with some comments on more general matter sources.

\subsection{Mutual information regularization}\label{subsec:mutual}

As discussed in \S \ref{sec:intro}, a difficulty in implementing the sum rule (\ref{fgf}) in QFT is that in general both sides are divergent. On the other hand, we just found that holographic regularization in terms of a cutoff at $z=z_{UV}$ makes the entanglement entropy and $\langle \Theta(x) \Theta(0)\rangle$ simultaneously well-defined, and provides a perfect match between such quantities in the holographic sum rule. In order to understand better this `nice' regulator, we now compare it with the result in terms of the mutual information, which gives a point-splitting regularization for the entanglement entropy.

Mutual information is a combination of entropies of three regions 
\be
I(A,B)=\mc S(A)+\mc S(B)-\mc S(A\cup B)\,,
\ee
for non intersecting $A$ and $B$.
Because the divergent terms are local and extensive on the entangling surface, they cancel in this combination, and mutual information
 is regularization independent in the continuum limit for any regions $A$ and $B$. It can be used as a regularization of entropy taking the limit 
 when the entangling surfaces of $A$ and $B$ are close to each other. This is analogous to framing regularization for Wilson loops. 
 In the present context we take as $A$ and $B$ two parallel planar entangling surfaces separated by a distance $l$. $\mc S(A\cup B)$ corresponds to the entropy of a thin strip of width $l$. In the holographic framework we then have
 \be
 I= \frac{1}{4 G^{(d+1)}} (2 A_{\text{plane}}- A_s)\,, \label{lpo}
 \ee 
 where $A_{\text{plane}}$ is the area of the minimal surface corresponding to a plane and $A_{s}$ the one corresponding to a thin strip. 
  
 We argue that the constant term in the entropy is the same as half the constant term (or logarithmic term) for the mutual information, showing that these terms are universal despite the possible presence of non analytic divergences. Essentially, the strip term does not correct these universal terms. The general argument is simple. For sufficiently small $l$, the strip minimal surface only tests the UV part of the bulk, where the metric can be expanded as the AdS metric plus corrections which are given by a series of powers in the coordinate $z$, starting with $z^{2(d-\Delta)}$. The calculation of the minimal surface and the area of the strip is perturbative in these corrections of the metric, and as a result the area is also given as a power series in the UV cutoff $\delta$ and the strip width $l$. The divergent terms in powers of $\delta$ must exactly cancel those of $2 A_{\text{plane}}$ in (\ref{lpo}) producing a finite mutual information. The rest of the strip contribution can be organized as a power series in $l$. For generic values of $\Delta$ the powers of $l$ are either smaller than zero, contributing to the divergent terms in the mutual information as a function of $l$, or positive powers, which can be neglected in the small $l$ limit. Then the constant term does not get modified from the one provided by $A_{\text{plane}}$. This, in contrast to the strip term, contains information on the whole RG running and the metric deep in the bulk. For some special values of $\Delta$ we could in principle get a $z^0$ term in the area of the strip. However, the area is some integral over $z$, and a zero power comes as a result of $\int dz/z$, giving a logarithmic term instead. In this particular case, the logarithmic term must come in a combination $\log(l/\delta)$ because the integral in $z$ runs from a UV cutoff $\delta$ to some maximal reach of the minimal surface in the bulk which is proportional to $l$. Again, the $\log(\delta)$ must be cancelled by the logarithmic term in $2 A_{\text{plane}}$. As a result, the logarithmic $\log(\delta)$ term in the entropy has exactly the same coefficient as the  $\log(l)$ term in the mutual information. In the presence of a logarithmic term, this coefficient is universal, while the constant term is not.

Let us make a simple calculation to illustrate this idea, expanding the metric near the boundary to the first subleading power and computing the strip entropy up to this order. Depending on the spacetime dimension and the particular powers 
 appearing in the metric expansion one should carry on the expansion to higher order terms. However, our point is that no corrections to the constant term appear in the strip term for generic values of the powers, and our calculation will be enough to illustrate this. A similar calculation was carried out in \cite{F}.        

\begin{figure}[h]
\begin{center}  
\includegraphics[scale=0.7]{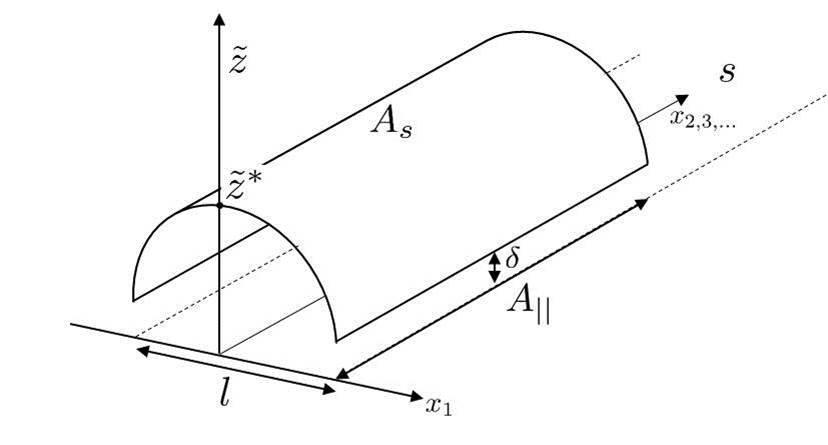}
\captionsetup{width=0.9\textwidth}
\caption{Strip geometric set-up: The strip $s$ is the region between the two planes  represented by two discontinuous lines. The planes extend along the $\left\lbrace x_2, x_3, ...\right\rbrace $ coordinates and are separated along the $x_1$ coordinate a distance $l$, the strip width. $\tilde{z}$ is the bulk radial coordinate and $\tilde{z}^{\ast}$ is the maximum reached by $A_s$, the bulk minimal-area-surface that is homologous to $s$. $\delta$ is a UV-cut-off and $A_\parallel$ is the area of the planes.  \label{fig}}
\end{center}  
\end{figure}

The dependence of $A_s$ on the width $l$ of the strip is obtained by solving
\begin{eqnarray}
A_s&=& 2 L_{UV}^{d-1} A_{\parallel} \frac{1}{\tilde{z}^{\ast (d-2)}} \int_{\delta/\tilde{z}^{\ast}}^1 d v\ \frac{1}{v^{d-1}}\frac{1}{\sqrt{f(\tilde{z}^{\ast} v )} \sqrt{1-v^{2(d-1)}}} \,,\label{A} \\
l&=&2 \tilde{z}^{\ast} \int_0^1 d v \ \frac{v^{d-1}}{\sqrt{f(\tilde{z}^{\ast} v )} \sqrt{1-v^{2(d-1)}}}\ .\label{l} 
\end{eqnarray}
Here  $\delta$ is an UV-cut-off,  $A_\parallel$ is the area of the planes defining the strip, and  $\tilde{z}^{\ast}$ is the maximum in the $\tilde{z}$ bulk radial coordinate reached by $A_s$, see figure \ref{fig}. $f(\tilde{z})$  defines the generic bulk metric\footnote{For convenience we have changed
coordinates $\frac{d\tilde{z}}{\sqrt{f(\tilde{z})}}=dz$ with respect to the $z$ coordinate used in previous sections. }
$$ds^2= \frac{L^2}{\tilde{z}^2}\left(dx^2 + \frac{d\tilde{z}^2}{f(\tilde{z})} \right) \,, $$ 
and describes the behavior of the $d$-dimensional boundary theory under the RG flow. \\

We solve for $A_s(l)$ in the limit $m \,l \ll 1$, were $m$ is the scale characterizing the leading relevant perturbation of the UV fixed point. In the bulk geometry, this corresponds to the limit where $A_s$ only probes the near AdS geometry given by
\begin{equation}
f(z)= 1 + (m \tilde{z})^{2\nu}+\hdots  \ ,          \label{pert} 
\end{equation}
with $\Delta = d- \nu < d$, the conformal dimension of the operator carrying the leading UV deformation. Specifically, we solve (\ref{A}--\ref{l}) at order $(m \tilde{z}^{\ast})^{2\nu}\sim (m l)^{2\nu}<< 1$. From (\ref{l}) and (\ref{pert}) we have
\begin{equation}\label{zpert} 
\tilde{z}^{\ast}=\frac{l}{2a}\left( 1 + \frac{b}{4^{\nu} a^{2 \nu +1}} (m l)^{2 \nu}+\hdots\right) \ ,
\end{equation}
with
\begin{eqnarray}
a&=&\int_0^1 d v \frac{v^{d-1}}{\sqrt{1-v^{2(d-1)}}}=\sqrt{\pi} \frac{\Gamma(\frac{d}{2(d-1)})}{\Gamma(\frac{1}{2(d-1)})}\,,\nonumber\\
b&=& \frac{1}{2} \int_0^1 d v \frac{v^{d-1+2 \nu}}{\sqrt{1-v^{2(d-1)}}}=\frac{\sqrt{\pi}}{2(1+2 \nu)}\frac{\Gamma(\frac{d+ 2\nu}{2(d-1)})}{\Gamma(\frac{1 + 2\nu}{2(d-1)})}\ . \nonumber 
\end{eqnarray}
At the same order, we get from (\ref{A})
\begin{eqnarray}\label{strip} 
&& A_s/(2 L^{d-1} A_\parallel)= \nonumber\\
&&=-\frac{a}{d-2}\frac{1}{\tilde{z}^{\ast(d-2)}}- \frac{1+ 2 \nu}{2-d+ 2 \nu} b \frac{(m \tilde{z}^{\ast})^{2 \nu}}{\tilde{z}^{\ast(d- 2)}}  + \frac{1}{d-2}\frac{1}{\delta^{d-2}} - \frac{1}{2}\frac{(m \delta)^{2\nu}}{(d-2-2 \nu) \delta^{d-2}}+\hdots\ \label{stri}\ \ \ \\
&&= -\frac{2^{d-2}}{d-2}\frac{a^{d-1}}{l^{d-2}} + \frac{d-1}{d-2-2\nu}(2 a)^{d-2-2 \nu} b \frac{(m l)^{2 \nu}}{l^{d-2}}  + \frac{1}{d-2}\frac{1}{\delta^{d-2}} - \frac{1}{2}\frac{(m \delta)^{2\nu}}{(d-2-2 \nu) \delta^{d-2}} +\hdots \, ,\ \ \ \ \nonumber 
\end{eqnarray}
where we used (\ref{zpert}) in the last line. When forming the mutual information $I_s$ 
the last two terms in (\ref{stri}) exactly cancel the UV-divergent terms contained in $A_{\text{plane}}$, which is $4G^{(d+1)}$ times the entanglement entropy of the half space. We then have
\be
4G^{(d+1)} I_s= \frac{2^{d-2}}{d-2}\frac{a^{d-1}}{l^{d-2}} - \frac{d-1}{d-2-2\nu}(2 a)^{d-2-2 \nu} b \frac{(m l)^{2 \nu}}{l^{d-2}}+\hdots 
\ee

We see that the strip entanglement entropy has a power series expansion determined by the powers appearing in the metric expansion. For generic powers it will not contain a constant term in limit $l \rightarrow 0$. Then, any constant term appearing in the mutual information of the strip comes entirely from the entanglement entropy of the half space.

\subsection{Holographic analysis of reflection positivity}\label{subsec:positivity}

The holographic formula (\ref{erase}) gives the two-point function for the stress-tensor trace in terms of the ratio $\delta A'/\delta A$ near $z=0$. This is in turn fixed by imposing regularity in the IR. From the field theory side, the two-point function has to be consistent with unitarity, and we would like to understand how this appears in the gravity side. We will first prove in general that unitarity of the large $N$ QFT requires stability of the classical gravitational action under bulk perturbations. We will then focus on the stress tensor correlator derived before, verifying that the NEC together with regularity of the solution give a unitary result.

Consider a local operator $\mc O(x)$ in QFT. The Euclidean correlation function in a unitary theory satisfies reflection positivity (RP)
\be
\int d^dx\,d^dy\, \alpha^*(\bar{x})\langle \mc O(x)\mc O(y)\rangle \alpha(y)\ge 0\,,\label{rp}
\ee 
where $\alpha(x)$ is any smooth test function with support in the upper half of Euclidean space $x^0>0$, and $\bar{x}=(-x^0,x^1,...,x^{d-1})$. Then $\alpha^*(\bar{x})$ has support on the lower plane. When the QFT has a holographic dual, the on shell Euclidean action in presence of a source $\phi^0$ at the AdS boundary reads
\be
S(\phi^0)=-\frac{1}{2}\int d^dx\, d^dy\, \phi^0(x) \langle \mc O(x)\mc O(y)\rangle\phi^0(y) + ...\,.\label{sds}
\ee
where we have omitted divergent terms that make this action positive. Note that (\ref{sds}) involves the correlators at coincident points while (\ref{rp}) does not.

We want to find the conditions that ensure the RP property in holographic models. In order to see this let us choose $\phi^0_1$ and $\phi^0_2$ to have support for $x^0>0$, and let $\phi^0_{1\bar{1}}(x)=\phi^0_1(x)+\phi^0_1(\bar{x})$,  $\phi^0_{2\bar{2}}(x)=\phi^0_2(x)+\phi^0_2(\bar{x})$, $\phi^0_{1\bar{2}}(x)=\phi^0_1(x)+\phi^0_2(\bar{x})$, and $\phi^0_{2\bar{1}}(x)=\phi^0_2(x)+\phi^0_1(\bar{x})$. We have 
\bea
S(\phi^0_{1\bar{2}})+S(\phi^0_{2\bar{1}})-S(\phi^0_{1\bar{1}})-S(\phi^0_{2\bar{2}})\hspace{6cm}\nonumber \\=\int d^dx\,d^dy\, (\phi^0_1(\bar{x})-\phi^0_2(\bar{x})) \langle \mc O(x)\mc O(y)\rangle (\phi^0_1(y)-\phi^0_2(y))\ge 0\label{rp1}
\eea
by reflection positivity. Then RP requires this particular inequality for the action as a function of the boundary conditions. 

In order to prove this relation consider the action $S(\phi^0_{1\bar{2}})$. This is the bulk action of a bulk field $\phi_{1\bar{2}}(x,z)$ which has boundary condition $\lim_{z\rightarrow 0}\phi_{1\bar{2}}(x,z)= \phi^0_{1\bar{2}}(x)$.  The bulk action is local, and we can write it as a sum of two terms, $S^+_{1\bar{2}}$ and $S^-_{1\bar{2}}$, corresponding to the actions for $x^0>0$ and $x^0<0$,
\be
S(\phi^0_{1\bar{2}})=S(\phi_{1\bar{2}})=S^+_{1\bar{2}}+S^-_{1\bar{2}}\,.\label{23}
\ee  
Analogously, we have 
\be
S(\phi^0_{2\bar{1}})=S(\phi_{2\bar{1}})=S^+_{2\bar{1}}+S^-_{2\bar{1}}\,.\label{32}
\ee 

By symmetry under Euclidean time reflection the time reflected solutions $\phi_{1\bar{2}}$ and $\phi_{2\bar{1}}$ coincide at $x^0=0$, i.e., $\phi_{1\bar{2}}(x^0=0,\vec{x},z)=\phi_{2\bar{1}}(x^0=0,\vec{x},z)$. Hence we can take a continuous bulk field given by $\psi_{1\bar{1}}=\theta(x^0)\phi_{1\bar{2}}+\theta(-x^0)\phi_{2\bar{1}}$ that has boundary condition $\phi^0_{1\bar{1}}$. This is not a solution of the equations of motion for these boundary conditions, and we expect that the action is minimized by the solution of the equations of motion $\phi_{1\bar{1}}$ with these same boundary conditions. Then we have
\be
S(\psi_{1\bar{1}})=S^+_{1\bar{2}}+S^-_{2\bar{1}}\ge S(\phi^0_{1\bar{1}})\,.\label{12}
\ee
Analogously, defining $\psi_{2\bar{2}}=\theta(x^0)\phi_{2\bar{1}}+\theta(-x^0)\phi_{1\bar{2}}$ we have
\be
S(\psi_{2\bar{2}})=S^+_{2\bar{1}}+S^-_{1\bar{2}}\ge S(\phi_{2\bar{2}})\,.\label{21}
\ee
Combining (\ref{23})--(\ref{21}) we get RP, equation (\ref{rp1}). This argument may fail for higher derivative Lagrangians, reflecting potential violations of unitarity in these theories.

It is interesting that RP is warranted by the stability of the bulk solution, or in other words, the fact that the bulk solution for a given boundary condition should be an absolute minimum of the action. This stability is expected to hold in physically motivated models, while fully proving it in detail for a specific case may be challenging. 

This proof of RP is similar to the proof of strong subadditivity of holographic entropy \cite{matt}, though details differ, i.e., the role of Euclidean time reflection symmetry (analogous to CPT symmetry in Minkowski space) in the present proof.   
For the case of Wilson loop operators whose holographic dual is given by minimal surfaces, or fields with large dimension such that the bulk solution for point like insertions at the boundary is given by geodesics, reflection positivity follows, in a completely analogous way, from the triangle inequality for the minimal area (or length) of the bulk geometric object \cite{pos}. 

Let us now turn to a more detailed discussion of unitarity for the stress tensor correlator.
In momentum representation, RP is equivalent to the positivity of the spectral density $\rho(m^2)$ in the spectral representation of the correlator of stress tensors \cite{cappelli}
\be
\langle \Theta(p)\Theta(-p)\rangle=\int_0^\infty dm\, \rho(m^2) \frac{p^4}{p^2+m^2}\,.
\ee
To make contact with (\ref{erase}) this expression is subject to subtraction of a polynomial expansion around $p^2=\infty$ to eliminate UV contact terms. 

The spectral density can be extracted from this expression as the imaginary part
\be
\rho(m^2)=\frac{1}{\pi m^{3}} \,\Im \langle \Theta(p)\Theta(-p)\rangle|_{p^2=-m^2-i \epsilon}\,.
\ee
This is insensitive to analytic terms, in particular to contact terms.\footnote{This is equivalent to computing the Wightman correlator in Minkowski space in momentum space. }
According to (\ref{eq:eomz}) we have to consider now the equation for negative $-p^2$
\be\label{pinvertido}
\frac{d}{dz} \left(a^{d-1}(z) \varepsilon(z) \frac{d \delta \tilde{A}}{dz} \right)+a^{d-1}(z)\varepsilon(z) p^2 \delta \tilde{A}(z)=0\,,
\ee
and compute [see (\ref{erase})]
\be\label{eq:rho1}
\rho(p^2)=\frac{d-1}{\pi p^{3} \kappa^2} \,a^{d-1}(z) \varepsilon(z)\,\Im\, \frac{\partial_z\delta \tilde{A}}{\delta \tilde{A}}\Big|_{z \to 0}\,.
\ee
Let us assume the NEC, such that $a^{d-1}(z) \varepsilon(z)>0$ everywhere except possibly at $z=0$. This implies that radial evolution for $\delta \t A(z)$ is regular.

Since the fluctuation $\delta \t A(x)$ is real, its Fourier components obey $\delta \t A_p^* = \delta \t A_{-p}$. The spectral density may then be rewritten as
\be\label{eq:rho2}
\rho(p^2)=\frac{d-1}{\pi p^{3} \kappa^2} \frac{1}{|\delta \t A_p|} \left \lbrace \frac{1}{2i}\,a^{d-1}(z) \varepsilon(z) \left(\delta \t A_p^* \partial_z \delta \t A_p- \delta \t A_p \partial_z \delta \t A_p^* \right) \right \rbrace\Big|_{z \to 0}\,.
\ee
Let us normalize the solution such that $|\delta \t A_p| \to 1$ as $z \to 0$. The spectral density is thus the flux of probability for $\delta \t A$ (interpreting the radial direction as time evolution). This flux is conserved by the equation of motion,
\be
\partial_z \left[ a^{d-1}(z) \varepsilon(z) \left(\delta \t A_p^* \partial_z \delta \t A_p- \delta \t A_p \partial_z \delta \t A_p^* \right)\right]=0\,.
\ee
As a result, the spectral density may be evaluated at any $z$.

Calculating the flux for sufficiently large $z$, where the expansion for $\delta \t A_{IR}$ in (\ref{eq:solAdS}) holds,\footnote{The analytic continuation from the euclidean solution gives $\delta \t A_p(z) \propto i(p z)^\alpha H_\alpha^{(1)}(p z)$.} obtains
\be
\Im(\delta \t A_p^* \partial_z \delta \t A_p)=\frac{2}{\pi}\,|D_1(p)|^2 p^{2\aI}z^{2\aI-1}\,.
\ee
Here $D_1(p)$ is the constant factor in the IR solution (\ref{eq:solAdS}) that is determined through the matching procedure in terms of the boundary source $h_0(p)$. Adding the dependence from the warp factor and $\varepsilon(z)$ in the far IR limit, we arrive to our final result for the spectral density
\be\label{eq:holodensity}
\rho(p^2)=\frac{2}{\pi^2}\,((\DI-d)L_{IR}^{(d-1)/2}\phi_{IR}^0)^2\,|D_1(p)|^2 p^{2\DI-d-3}\,.
\ee
Here $\phi^0_{IR}$ determines the approach to the IR AdS solution in (\ref{eq:phiIR}); in QFT language, $(\DI-d)L_{IR}^{(d-1)/2}\phi_{IR}^0$ is the $\beta$ function for the leading irrelevant operator that dominates the flow towards the IR fixed point.

The spectral density (\ref{eq:holodensity}) is positive definite, so this establishes the RP property of the stress tensor correlator in the holographic model. Therefore, the NEC together with regularity in the IR ensure that the two-point function of $\Theta(x)$ is unitary. Of course, a further assumption is that $\rho(p^2)$ exists and is finite as the limit $z \to 0$ is taken in (\ref{eq:rho2}). 

Given this, it is interesting to ask how the unitarity bound $\DU>(d-2)/2$ could be seen in the stress tensor correlator.
In this work we have restricted to $\DU >d/2$, namely the standard quantization; this is stronger than the unitarity bound, which hence does not appear as a further restriction on the gravity solution. We could naively (and, as it turns out, incorrectly) extrapolate our formulas to $\DU<d/2$, finding a divergent answer for the spectral density from the limit $z \to 0$. However, this is not correct because the alternate quantization $\DU<d/2$ requires a different boundary value problem in terms of Neumann boundary conditions~\cite{Klebanov:1999tb}. It will be interesting to extend our analysis to RG flows with $\DU<d/2$, something that we hope to address in the future.

\subsection{Structure of the $p^d$ term}\label{subsec:pd}

Using (\ref{eq:thetafinal}) we may also investigate the next terms in the expansion for low momentum of the correlation function of $\Theta$. The first non integer power of $p$ comes from the term $\frac{A_1}{A_2} \propto p^{2\Delta_{IR}-d}$ in (\ref{eq:thetafinal}). Since $\Delta_{IR}>d$ the expansion is in terms of integer powers of $p^2$ up to $p^d$. 
In particular in even dimensions the interesting dimensionless quantity
\be
\int d^d x \,x^d \langle \Theta(x) \Theta(0) \rangle=(-1)^{d/2} (\nabla_p^2)^{d/2} \langle \Theta(p)\Theta(-p)\rangle \vert_{p=0}
\ee
is given by purely geometric integrals in holographic theories. This quantity has been analyzed in the past in connection to RG irreversibility \cite{anselmi}. 

The term proportional to $p^d$ is determined by expanding
\be
\langle \Theta(p)\Theta(-p)\rangle \approx \frac{d-1}{\kappa^2}\,\frac{1}{f_0'}\frac{p^2 g_1'+p^4 g_2'+\ldots}{1+p^2 g_1+p^4 g_2+\ldots}\Big|_{z=\zU}\,.
\ee
For instance in $d=4$ the coefficient of $p^4$ is 
\be
 \frac{3}{\kappa^2}\frac{1}{f_0'} (g_2'-g_1 g_1')= \frac{3}{\kappa^2} \int_{z_{UV}}^{z_{IR}} d y_1\, \frac{1}{a^3(y_1)\varepsilon(y_1)} \left(\int_{z_{IR}}^{y_1} dy_2\, a^3(y_2)\varepsilon(y_2)\right)^2\,. \label{coefi}
\ee
This is positive and UV and IR finite for generic flows. However, this dimensionless quantity is not a boundary term. Hence it depends on the details of the flow and does not reduces to 
a difference of anomalies between fixed points in general. In \cite{anselmi} it is claimed that this is proportional to
the change of the $a$ anomaly between fixed points for marginally relevant flows. A similar statement is made in \cite{Kaplan:2014dia} in the limit of
 ``slow roll'' solution for the domain wall. We were not able to find evidence in support of these claims from (\ref{coefi}), although it would be interesting to understand, in our framework, the simplifications entailed by nearly marginal flows.

\subsection{Comments on more general matter sectors}\label{subsec:gen-matter}

So far we have studied RG flows that are described holographically in terms of a single scalar field with canonical kinetic term and a potential $V(\phi)$ with two $AdS$ critical points. Nevertheless, the result for the stress tensor two-point function should hold more generally, for instance in the presence of multiple scalars or with small higher derivative terms --as long as unitarity is maintained. Here we will comment briefly on some of the new issues that arise for more general matter sectors, and suggest a possible method of analysis which we hope to apply in future work.

Let us focus for simplicity on the case of multiple scalar fields, corresponding to turning on many relevant deformations of the UV fixed point. The perturbative expansion around CFT$_{UV}$ will be dominated by the most relevant deformation, but as the flow proceeds we expect a rather complicated dynamics involving the other deformations as well. The approach to CFT$_{IR}$ will be dominated by the leading irrelevant operator. Holographically, we have a domain wall describing a trajectory in field space that interpolates between a local maximum and a minimum of $V(\phi)$. The goal is to compute the $\langle \Theta(x) \Theta(0) \rangle$ in this background.

Einstein's equations are similar to (\ref{eq:GReqs}), after including the total kinetic and potential energy contributions of all the scalars. On the other hand, the scalar field equations of motion are now independent from the gravitational equations (except for one). Let us then analyze these new equations at the linearized level, as needed for the stress-tensor two point function. The starting point is the scalar equation in the presence of the lapse and shift functions $N$ and $N_\mu$:
\be
\partial_r \left(\sqrt{h}N N^{-2} (\dot \phi_i - N^\mu \partial_\mu \phi_i) \right)+ \partial_\mu \left(\sqrt{h}N[-N^{-2}N^\mu(\dot \phi_i - N^\nu \partial_\nu \phi_i)+ \partial^\mu \phi] \right)=\sqrt{h}N \frac{\partial V}{\partial \phi_i}\,.
\ee
Linearizing this equation for $N(x,r) = 1+ \delta N(x,r)$, $N_\mu(x,r) = \delta N_\mu(x,r)$, $\phi_i(x,r)=\phi_i(r) + \delta \phi_i(x,r)$ obtains
\be\label{eq:multiple1}
\delta \ddot \phi_i + (d \,\dot {\delta A} - \dot {\delta N}) \dot \phi_i + d \dot A \,\delta \dot \phi_i - \dot \phi_i \partial_\mu \delta N^\mu+ \Box \delta \phi_i-2 \frac{\partial V}{\partial \phi_i}\,\delta N - \frac{\partial^2 V}{\partial \phi_i \partial \phi_j} \delta \phi_j=0\,.
\ee

The main issue with extending our approach of \S \ref{sec:proof} to this case
is that it is no longer possible to choose a gauge where all scalar fluctuations vanish. To see this, proceed by contradiction and assume that $\delta \phi_i=0$; $\delta N$ and $\delta N_\mu$ are the same as before, and then (\ref{eq:multiple1}) evaluated on $\delta \phi_i=0$ gives
\be
\left(\ddot \phi_i -\dot \phi_i \frac{\sum_j \dot \phi_j \ddot \phi_j}{\sum_j \dot \phi_j^2} \right)\dot{\delta A}=0\,.
\ee
This is trivial for a single scalar field --showing that the gauge $\delta \phi=0$ is consistent-- but the equation cannot be satisfied for multiple fields. We conclude that with many scalar fields a metric fluctuation $\delta A$ will source fluctuations $\delta \phi_i$, and these will contribute to the stress-tensor two-point function.

In order to incorporate these and other more general effects, it seems useful to think in terms of an arbitrary matter energy-momentum tensor $T_{MN}$ in the bulk. The linearized Einstein's equations will then include density, pressure, momentum and stress fluctuations from $T_{MN}$. A natural extension of \S \ref{sec:proof} to these general `fluids' is to choose the uniform density gauge $\delta \rho=0$. In fact, a similar situation arises in cosmology with multiple inflatons; see e.g.~\cite{Baumann:2009ds} for a recent review. We expect that by imposing the NEC on $T_{MN}$, together with the positivity constraint of \S \ref{subsec:positivity}, the holographic sum rule will hold.
We hope to return to this point in the future.

\section{Conclusions and future directions}\label{sec:concl}

In this work we have calculated the stress tensor two-point function $\langle \Theta(x) \Theta(0)\rangle$ for holographic renormalization group flows between pairs of conformal field theories. Imposing regularity in the bulk interior and matching onto the UV fluctuation, we obtained the two-point function in a series expansion at small momenta, Eq.~(\ref{eq:thetafinal}). This result is valid for general scalar potentials, with the coefficients of the series determined in terms of the background warp factor and its derivatives. We showed that the leading $p^2$ term gives the change in the central charge for $d=2$, while in $d>2$ it reproduces the entanglement entropy for a planar surface. This provides a holographic realization for the result in~\cite{area}. Finally, we showed in general that reflection positivity of the boundary QFT requires stability of the gravitational action under bulk perturbations. For the class of models considered here, this is implied by the NEC and regularity of the solution.

Let us end by summarizing some future directions of research motivated by these results. First, it would be very interesting to extend holographic RG flows and the calculation of the stress tensor two-point function to more general matter sectors. As discussed briefly in \S\ref{subsec:gen-matter}, it may prove useful to formulate the problem directly in terms of perturbations of the energy momentum tensor, as done in cosmology. Even at the level of a single two-derivative scalar field, there remains the question of flows with the alternate quantization, and how the transition to the standard quantization occurs due to the domain wall.

Another direction involves studying cases with spontaneous conformal symmetry breaking. This may be related to a different issue worth studying: the role of improvement terms in the bulk and how they modify the stress tensor correlator and the entanglement entropy. The holographic sum rule may also have implications for inflationary models connecting de Sitter solutions. Finally, it would be interesting to incorporate corrections to both sides of the sum rule, both from $1/N$ and $g_s$ effects.

\section*{Acknowledgments}
We thank V. Hubeny, J. Kaplan, R. Myers, M Rangamani, M. Smolkin, R. Trinchero and J. Wang for interesting discussions. This work was supported by CONICET, ANCyT, Universidad Nacional de Cuyo, and CNEA, Argentina.



\bibliographystyle{JHEP}
\renewcommand{\refname}{Bibliography}
\addcontentsline{toc}{section}{Bibliography}
\providecommand{\href}[2]{#2}\begingroup\raggedright

\end{document}